\documentstyle[11pt,aaspp4]{article}
\newcommand{\etal}{{\it et al.\ }}
\newcommand{\tP}{\tilde{P}}

\newcommand{\avg}[1]{\left\langle{#1}\right\rangle}

\renewcommand{\bar}{\overline }

\newcommand{\nb}{\bar{N}}
\newcommand{\xiav}{\bar{\xi}}
\newcommand{\mpc}{h^{-1}{\rm Mpc}}
\newcommand{\kpc}{h^{-1}{\rm kpc}}
\newcommand{\neff}{n_{\mbox{\it eff}}}

\begin{document}

\Large 

\title{ $N$-point correlations in CDM and $\Omega$CDM Simulations
       }
\normalsize 
\vskip 0.5cm 
\author{Istv\'an Szapudi$^1$, Thomas Quinn$^2$, Joachim Stadel$^2$,
        and George Lake$^2$} 
\vskip 1cm 
\affil{$^1$University of Durham, Department of Physics 
 South Road,  Durham DH1 3LE, United Kingdom}
\affil{$^2$ Department of Astronomy, University of Washington,
  Seattle, WA 98195-1580}
\vskip 0.3cm

\vskip 3cm 
\centerline{\bf Abstract} 

\small 
Higher order statistics are investigated in ($\Omega$)CDM universes
by analyzing $500\mpc$ high resolution tree $N$-body
simulations with both $\Omega = 1$, and $\Omega < 1$. 
The amplitudes of the $N$-point correlation functions
are calculated from moments of counts-in-cells determined
by a pair of new algorithms  especially developed for 
large simulations. This approach enables massive oversampling
with $\simeq 10^{9-14}$ cells for accurate determination of
factorial moments from up to $47$ million particles in the
scale range of $8 \kpc - 125\mpc$.
Thorough investigation
shows that there are three scale ranges in the simulations:
$\ge 8\mpc$, weakly non-linear regime, where perturbation
theory applies with utmost precision, $1\mpc - 8\mpc$, the non-linear
plateau, and finally $\le 1\mpc$, a regime where dynamical 
discreteness effects dominate the higher order statistics.
In the physically relevant range of $1\mpc-125\mpc$ the results
i) confirm the validity of perturbation theory in the
weakly non-linear regime, ii) establish
the existence of a plateau in the highly non-linear 
regime similar to the one observed in
scale free simulations iii) show extended perturbation theory
to be an excellent approximation for the non-linear
regime iv) find the time dependence
of the $S_N$'s to be negligible in both regimes v)
in comparison with similar measurements in the EDSGC
survey, strongly support $\Omega < 1$ with no biasing vi) show that
the formulae of Szapudi \& Colombi (1996) provide a good approximation
for errors on higher order statistics measured in $N$-body 
simulations.

\normalsize
\vskip 0.5cm
{\bf \noindent keywords} large scale structure of the universe -- galaxies: clustering
-- methods: numerical -- methods: statistical

\section{Introduction}

According to popular theories of structure formation,
the distribution of mass in the universe grows by gravity
from initially Gaussian fluctuations.
The resulting distribution is described in a statistical way, most 
importantly via two-point and higher order correlation functions,
which can be studied theoretically using analytical
methods,  or numerical experiments. Although the comparison
of the results with observations is somewhat complicated by
the fact that galaxies do not necessarily trace mass (biasing),
the manyfold information contained in the higher order correlations
in principle enables the separation of gravitational amplification from other
processes (e.g.,
\cite{f94,mvh97,scffhm98,fg98,ssf98,sz98}).

Following the pioneering work of Peebles and collaborators
(e.g., \cite{fp78,peebles80} and references therein),
perturbation theory became the prime analytical tool to study
higher order correlation functions. The Euler equations for
a gravitating fluid are expanded around small fluctuations to predict
the amplitudes of the correlation functions 
at weakly non-linear scales. In contrast, $N$-body simulations
calculate the gravitational amplification directly; thus, up to
numerical accuracy, they follow the full non-linear evolution.
Simulations not only yield beautiful agreement with perturbation theory
at large scales, but also penetrate the highly non-linear evolution
of smaller scales. These scales are especially important, since,
except for the largest galaxy catalogs, most observations are
performed at small or intermediate scales. The method of
moments of counts in cells is especially useful for comparison,
since the moments were calculated in the
framework of perturbation theory (e.g.,
\cite{peebles80,jbc93,bern92,bern94,bern95}), 
measured in $N$-body simulations (e.g., \cite{bsd91,bh92,bge95,gb95,cbh96}),
and galaxy catalogs as well 
(e.g., 
\cite{peebles80,gaz92,ssb92,mss92,bouchet93,gaz94,sdes95,smn96,ks97,ss97}).

While the simplest version of SCDM initial conditions appears
to be excluded by observations of the variance as measured by
the Cosmic Microwave Background, cluster abundances, pair-wise
velocities, and galaxy clustering, it is the qualitatively most
successful theory, to which every other theory is measured. In this work
large, high resolution CDM simulations are used in an attempt
to understand clustering with unprecedented errors
in a large dynamic range. Motivated by observations, 
a low density variant of CDM ($\Omega$CDM) is investigated as well, since
it is one of the most viable alternatives at present. 

Moments of counts in cells are used to quantify 
higher order clustering in the simulations.
Similar previous measurements are improved upon in several ways:
a large $500\mpc$ box size is used to diminish finite volume effects,
i.e. the error on the measurement from fluctuations of the
universe on scales larger than the box size; $47\times 10^6$ particles
are used for a large dynamic range; a pair of new methods are
employed calculate counts in cells,
which are especially designed for large simulations and to minimize
the measurement errors; for quantitative assessment of the
accuracy a strict theoretical error analysis is
performed according the formalism of
Szapudi \& Colombi (1996 hereafter \cite{sc96}),
and Szapudi, Colombi, \& Bernardeau (1998, hereafter \cite{scb98}). 
Because of the above properties the measurements 
are relevant to study both the highly and mildly non-linear regimes
as well as the transition between them. Special care is taken
to determine the scales of reliability, and appropriate tests are
done to estimate the artificial two-body relaxation effects,
which appear to be the limiting factor at small scales.

The organization of the paper is as follows.
The next $\S $ outlines the method of counts in cells
as used here. $\S 3$ describes the simulations and
establishes the scales of reliability. $\S 4$ 
presents the measurements of the cumulants in the various
simulations. $\S 5$  discusses findings in terms of
perturbation theory (PT) and extended perturbation theory (EPT)
providing an efficient framework to compress the results 
and facilitating the comparison with observational data
from the EDSGC survey. The Appendix contains the
definition of the pair of algorithms used to calculate
counts in cells.

\section{Method}

A substantially improved version
of the counts in cells method is used in this work.\footnote{
Note that in an $N$-body simulation edge effects are
eliminated by the periodic boundary conditions, therefore the
edge corrected estimator of \cite{ss98} is not necessary.
}  It consists of calculating
the amplitudes of higher order correlation functions in 
a sequence of three consecutive steps:
estimation of the probability distribution,
calculation of the factorial moments, and extraction of the
normalized, averaged amplitudes of the $N$-point correlation functions,
the $S_N$'s. 
The relevant definitions and theory is briefly summarized below,
while \cite{smn96}, and references therein can be consulted for more
details. 

Let $P_N$ is the probability that a randomly thrown cell in the
simulation contains $N$ particles, 
with implicit dependence on the cell
size $\ell$. The estimator for this is the frequency distribution
\begin{equation}
  \tP_N  = {1 \over C}\sum_{i=1}^C \delta(N_i = N),
\end{equation}
where $C$ is the number of cells thrown and $N_i$ is the
number of objects in cell $i$. It is desirable to use
as many cells as possible, since for large $C$ 
the measurement errors associated with the finite number of cells
behave as $C^{-1}$ (\cite{sc96}). Here the main improvement
over the more traditional approach is the pair of algorithms
described in the Appendix, which enable us to use $C \simeq 10^{9-14}$
even in these large simulations.

The factorial moments (see e.g. Szapudi \& Szalay 1993)
may be obtained from the probability distribution using
\begin{equation}
  F_k = \sum P_N (N)_k,
\end{equation}
where $(N)_k =  N(N-1)..(N-k+1)$ is the $k$-th falling factorial of $N$.
The $F_k$'s directly estimate the moments of the hypothetical continuum
random field which is Poisson sampled by the simulation particles.
This is the most accurate and efficient way of subtracting shot noise,
which becomes important on small scales. Note, that for estimation
purposes, the estimator of the probability distribution is substituted
in the above equation, i.e. $\tP_N \rightarrow P_N$.

The average of the $N$-point angular correlation functions on a scale $\ell$
is defined by
\begin{equation}
  \xiav_N (\ell)=V^{-N}\int dV_1\ldots dV_N \xi_N(r_1,\ldots,r_N),
\end{equation}
where $\xi_N$ is the $N$-point correlation function in the simulations
and $V$ is the volume of a cell.
We define $S_N$ in the usual way,
\begin{equation}
   S_N = \frac{\xiav_N}{\xiav_2^{N-1}}.
\end{equation}
The factorial moments have an especially simple relation to
the $S_N$'s through the recursion relation (Szapudi \& Szalay 1993),
which is quoted for completeness:
\begin{equation}
  S_k = \frac{F_k\xiav_2}{N_c^k}-\frac{1}{k}\sum_{q=1}^{k-1}
        \frac{(k-q)S_{k-q}F_q{k \choose q}}{N_c^q},
\end{equation}
where $N_c = \avg{N} \xiav_2$.

The most critical and CPU intensive component of the above procedure is 
the calculation of counts in cells with appropriate oversampling.
While there exists an algorithm for infinitely  oversampling by \cite{sz97}, 
it would be impractical for $47$ million particles in three
dimensions. Therefore, a new approach was developed especially
for large simulations: the resulting
pair of algorithms for smaller and larger scales
have substantial overlap at intermediate scales suitable for testing. 
They are detailed in the Appendix.  With a modest $6-8$ hours
of CPU investment, these algorithms can achieve $C \simeq 10^{9-14}$ sampling
cells simultaneously at a hierarchy of scales between $1/65536-1/4$
of the simulation box size. 

\section{Measurements}

\subsection{Simulations}

\begin{table}
\begin{center}
\begin{tabular}{cccccccc}
\hline
Label & $\Omega$ & $H_0$ (km/s/Mpc) & $\sigma_8$ & $L_{box}$ ($\mpc$)&
N & $\epsilon$ ($\kpc$) & $l_{50}$ ($\mpc$)\\
\hline\\
$I_a$   & 1.0    & 50 & .5 & 500.0 & $4.7 \times 10^7$ & 50.0 & 4 \\
$I_b$   & 1.0    & 50 & .7 & 500.0 & $4.7 \times 10^7$ & 50.0 & 2\\
$I_c$   & 1.0    & 50 & 1.0 & 500.0 & $4.7 \times 10^7$ & 50.0 & 1\\
$II_a$   & 0.5   & 75  & .74 & 500.0 & $4.7 \times 10^7$ & 50.0 & 3\\
$II_b$   & 0.4   & 75  & .88 & 500.0 & $4.7 \times 10^7$ & 50.0 & 1.3\\
$II_c$   & 0.3   & 75  & 1.0 & 500.0 & $4.7 \times 10^7$ & 50.0 & $\le  1$\\
$i_{c0}$   & 1.0   & 50  & 1.0 & 500.0 & $3.0 \times 10^6$ & 160.0 & 13\\
$i_{c1}\ldots i_{c5}$   & 1.0   & 50  & 1.0 & 200.0 & $3.0 \times 10^6$ & 50.0 & 1\\
\hline
\end{tabular}
\end{center}
\caption{The simulations analyzed are summarized in this table.
$\Omega$ is the density parameter, $H_0$ is the Hubble constant,
$\sigma_8$ is the RMS density
fluctuation in $8\mpc$ spheres, $L_{box}$ is the size of the periodic box, N
is the number of particles, and $\epsilon$ is the force softening
length. Note
that $i_{c1} \ldots i_{c5}$ are five different realizations of the
same initial conditions.}
\label{simtab}
\end{table}

The characteristics of the simulations used are summarized in table
\ref{simtab}.  The box size, particle number, and force softening of
the large simulations were
chosen to model the formation of galaxy clusters in a volume of the
universe comparable to that to be surveyed by the Sloan Digital Sky
Survey (SDSS, \cite{gk93})  All simulations were computed using PKDGRAV,
(Stadel and Quinn, in preparation) a scalable parallel treecode with
periodic boundary conditions.  Accurate forces were maintained by
using a cell opening angle of $\theta = 0.8$ for $z < 2$ and $\theta =
.6$ for $z > 2$, and expanding the
potentials of cells to hexadecapole order.  Timesteps were constrained
to $\Delta t < 0.3 (\epsilon/v_{max})$, where $\epsilon$ is the
softening length and $v_{max}$ is the approximate maximum speed.  A
cubic spline softening kernel was used.  The simulations were started
at $z = 49$ for the $\sigma_8 = 1$ models, thus the
transients from initial conditions should be negligible (\cite{scoc98}).   
The same simulations were also used by \cite{gov98} to explore the
properties of galaxy clusters.
Note that $I_a$, $I_b$, and $I_c$ are
the same simulation at different output times.  $II_a$, $II_b$, and $II_c$ are
likewise.  Simulations $i_{c1}\ldots i_{c5}$ are an ensemble of
simulations with different realizations of the same initial power spectrum.

\subsection{Scales of Reliability}

Since both algorithms in the Appendix employ powers of $2$,
initially the scale range of $2^{-16} \ldots 2^{-2}$ times box size
was used for calculating counts in cells, corresponding
to $7.63\kpc-125\mpc$. The lower scale is smaller than
the softening length used for force calculation. With
our algorithm A2 for small scales we could obtain almost
arbitrarily small scales for free. The upper scale
still contains $256$ non-overlapping volumes, sufficient for
a distribution not far from Gaussian. Figure 1 shows the 
the counts in cells distribution for simulation $I_c$. The curves
from right to left correspond to scales of $1/4, 1/8, \ldots, 1/65536$ L
(the box size). Algorithm A1 was used for scales down to $1/512$ L, 
and A2 for smaller scales. 
A1 uses a fixed number of cells $C \simeq 1.1\ 10^{9}$
for all scales, while A2 increases $C$ $8$-fold at each step towards smaller
scales after starting with the above value. 
This is reflected on Figure 1 by the lowest possible value
$P_N$ can take. The tail of the distribution still
shows some wavering which could be smoothed out with even higher
oversampling. The next $\S$, however, will show that the resulting
measurement errors are much smaller than the theoretical 
variance of the simulations, thus the sampling is
sufficient. The high degree of oversampling was made possible only
by the algorithms of the Appendix specifically developed
for this purpose. 

The upper panel of
Figure 2 shows the variance, or the average two point correlation
function over a cell,  as calculated from the first two factorial moments
of simulation $I_c$. The solid curve is algorithm A1, while the joining
dashed line shows algorithm A2.  The triangles and squares display
the expected variance in cubic windows obtained by integrating the linear and
non-linear power spectrum, respectively (\cite{pd96}; the non-linear
$P(k)$ was
provided by Carlton Baugh, private communication). The fitting formula
for the non-linear power spectrum provides a good approximation, the
largest discrepancy being roughly 30\% on the smallest scales.
The lower resolution simulation, discussed later, and represented 
with dotted line and the figure, is in even better agreement with
the fitting formula. While providing a more accurate fitting formula
for high resolution simulations could be a topic of further investigation, 
this work concentrates on higher order statistics, the second order 
moment is only shown as a test.

The higher order factorial moments were calculated as well
from the counts in cells
according $\S 2$. The resulting $S_N$'s up to 9th order are displayed
on Figure 3. Again, 
the solid and and the continuing dash lines are the results from the
two algorithms of the Appendix for small and large scales.
Note the excellent agreement in the overlap, despite the fact
that the sampling is somewhat different because of the random 
shifts employed in A2. 

Qualitatively one can distinguish three regimes on Figure 3.
The dot-dash lines display theoretical predictions from
perturbation theory (\cite{jbc93,bern94}) up to sixth order.
The agreement is excellent
from scales upward of $8\mpc$, in the weakly non-linear
regime. Between $1-8\mpc$, in the highly non-linear regime, the $S_N$'s
are higher than the perturbation theory prediction because
of enhanced non-linear effects. They constitute a shallow
plateau in agreement with previous results from scale invariant
simulations (\cite{cbh96}). Finally downward from $1\mpc$ there seems to be
a third regime with a steeper rise. As illustrated next, this is
caused by artificial particle discreteness effects.

The lower panel of Figure 2. plots $N_c = \xiav \nb$,
the number of particles in a typical cluster, as a function of scale. 
$N_c$ indicates
how well the simulation represents the fluid limit.
For small values the dynamics in  typical clusters
is artificially dominated by particle discreteness effects,
a dynamical shot noise (\cite{cbh96}). Such
effects do not represent real physics since particles in the
simulation should follow the dynamics 
of the underlying smooth field. Indeed, at scales smaller than
$1\mpc$ $N_c$ becomes fairly small, which is likely to explain the sharp
rise in the $S_N$'s. 

To test this idea several auxiliary CDM simulations were run
with $3$ million particles, and the $S_N$'s were measured.
Simulation $i_{c0}$ had the same initial conditions and
box size as the main simulation, therefore the shot noise
is expected to turn up the $S_N$'s on larger scales,
if the above explanation is correct. According to the dotted
line on the lower panel of Figure 2, which displays $N_c$ for
$i_{c0}$, the break is expected to happen at around $10-15\mpc$, if
the suspected scaling with $N_c$ is correct. Indeed, this seems
to be the case for the dotted lines on Figure 3, supporting
the role of particle discreteness in the artificial increase of the
higher order cumulants.

For another test a  set of $5$ simulations were run, $i_{c1}\ldots i_{c5}$.
They had smaller box size $200^3\mpc^3$, to keep the 
average number of particles the same as the original simulation.
The initial conditions were independently generated for each 
realization. 
The ensemble average of these
simulations is expected to yield the same results upward from
$1\mpc$ as the original simulation, perhaps with less accuracy
because of the enhanced cosmic error caused by the smaller volume
(\cite{sc96}).
These measurements are displayed as triangles on Figure 3. The 
errorbars were calculated by estimating the dispersion of the
five simulations $i_{c1}-i_{c5}$; only the upper errorbar
is displayed for clarity.
The results  are in excellent
agreement with the expectations, further supporting the
idea that the third regime
at small scales is a sign of dynamical discreteness effects.


\cite{cbh96} found that if $l_c$ is
defined by $N_c(l_c) = 1$, a sufficient condition
for the fluid limit is $l \ge 1.5 l_c$. The location of
the break in the curves suggest a slightly
more conservative limit such that
$ N_c(l_{50}) = 50$.
This somewhat ambiguous prescription depends on the details
of the simulation and the desired precision of the agreement 
between the fluid limit and the measurements at each order. 
Our choice corresponds
to $1\mpc$ as the scale of reliability.
Note that the accuracy depends on the order,
deteriorating towards the higher moments. It seems more
logical to relax the required precision towards higher order
than to define a set of  scales of reliability
becoming larger with higher order. This somewhat arbitrary but
natural choice of $1\mpc$ is adopted for the measurements
performed in the rest of the simulations, but $l_{50}$ is
given for reference in Table 1 for each output.
Note also, that for the two-point correlation function only,
a smaller $N_c$, and a correspondingly smaller
scale,  is sufficient (see e.g., \cite{jain97}).

\section{Results}

According to the previous reasoning, it is meaningful to
extract higher order correlations down to $\simeq 1\mpc$ only, thus 
only algorithm A1 was sufficient for the rest of the measurements.
Although the force resolution would suggest a lower threshold, 
as detailed above, particle discreteness effects raise the scale of
reliability. Six outputs of two high resolution
CDM simulations summarized in Table 1 ($I_a,I_b,I_c,II_a,II_b,II_c,$)
were used to measure counts in cells on scales between $1-125\mpc$.

\begin{table}
\begin{center}
\begin{tabular}{ccccccc}
\hline
$l$ ($\mpc$) & $I_a$ & $I_b$ & $I_c$ & $II_a$ & $II_b$ & $II_c$ \\
\hline\\
0.976  &   7.98 & 6.45 & 5.73  & 10.22 & 8.43 & 7.94\\
1.953  &   6.41 & 5.57 & 5.00  & 8.49  & 7.60 & 7.05\\
3.9 &      4.61 & 4.55 & 4.38  & 6.09  & 6.24 & 6.08\\
7.81 &     3.47 & 3.54 & 3.56  & 4.30  & 4.54 & 4.78\\
15.625 &   2.87 & 2.89 & 2.93  & 3.38  & 3.51 & 3.68\\
31.25 &    2.43 & 2.47 & 2.52  & 2.87  & 2.94 & 2.99\\
62.5 &     1.87 & 1.92 & 1.96  & 2.20  & 2.25 & 2.29\\
125 &      3.34 & 2.40 & 2.07  & 2.07  & 2.03 & 2.00\\
\hline
\end{tabular}
\end{center}
\caption{ The measurements of $S_3$ in the different
simulations are tabulated. $l$ is the size of the cubical
window in which counts in cells where measured.
 The properties of simulations can be found in Table 1. }
\label{s3tab}
\end{table}

\begin{table}
\begin{center}
\begin{tabular}{ccccccc}
\hline
$l$ ($\mpc$) & $I_a$ & $I_b$ & $I_c$ & $II_a$ & $II_b$ & $II_c$ \\
\hline\\
0.976  &  121.1   & 74.8  & 58.1  & 222.7 & 133.6  & 121.5 \\
1.953  &  80.0    & 56.7  & 44.7  & 154.4 & 115.2  & 96.5  \\
3.9 &     41.4    & 38.7  & 34.7  & 80.7  & 80.8   & 74.1  \\
7.81 &    21.5    & 22.5  & 21.8  & 36.9  & 40.4   & 46.7  \\
15.625 &  13.4    & 13.6  & 14.2  & 20.1  & 22.5   & 25.7  \\
31.25 &   9.1     & 9.6   & 10.2  & 13.8  & 14.4   & 14.8  \\
\hline
\end{tabular}
\end{center}
\caption{ Same as Table 2. for $S_4$}
\label{s4tab}
\end{table}

The measurements of the $S_N$'s are displayed in Figures 4 with
solid lines, which is the main result of this paper.
The $\Omega = 1$, and $\Omega < 1$
simulations are displayed on the left, and right hand side, 
respectively.
For reference, the measured $S_3$ and $S_4$ are given
in Tables 2 and 3 as well.
The weakly non-linear regime on large scales is
distinguished from the non-linear plateau at small scales
in all cases.
The behavior of the higher order moments is qualitatively  similar
to scale invariant simulations (\cite{cbh96}).
Perturbation theory predicts that the $S_N$'s are independent
of the output times. This appears to be a good approximation 
even in the highly non-linear regime, especially down
to scales of $l_{50}$. For instance for the $\Omega = 1$ simulations
on $4\mpc$ ($l_{50}$  for $I_a$)  $S_3$ changes only about 5\%, which
is the same order as the errors.  Even $S_{10}$ is constant
within a factor of 2-3, i.e. the higher order moments are constant
within the errors (see next subsection). The decreasing 
trend on small scales can be explained by contamination effects
from particle discreteness.
As the scale of reliability moves to the left 
for the more relaxed, later simulations, the $S_N$'s decrease slightly.
These initial observations will be refined by comparing with
the predictions of PT and EPT
in the next section, after the error budget is detailed in
the next subsection. 

\subsection{Errors}

According to \cite{sc96}, the errors on the previous results can be classified
into measurement errors and cosmic errors.

The measurement errors arise from a finite
number of cells, $C$, being used to estimate the distribution of
counts in cells. The appropriate expression for the
error generating function is (see \cite{sc96} for details):
\begin{equation}
  E(x,y) = \frac{1}{C}\left[P(x y) - P(x)P(y)\right].
\end{equation}
The expansion of this equation yields the measurement error
in the $N$-th moment, which depends on the $2N$-th moment.
If $C \rightarrow \infty$, the contribution approaches $0$
as expected. We used this equation self-consistently to
obtain errors up to 5th order (since 10th order moments were
measured). The measurement error is largest at the smallest
scales. Figure 5 shows the relative
measurement error as a function of order for the $1\mpc$ scale
for $I_c$. Since there is a convex curvature on the graph,
the continuing dotted line is a conservative overestimation
of the errors for the orders $N > 5$.
This suggest that even at 10th order
the measurement errors contribute less than
10\%, thus a further increase in the sampling is not required.
This finding is true for the other simulations as well.

Cosmic errors are an inherent property of the simulations
and cannot be improved upon, except by using a larger volume
or an ensemble of realizations. This type of error can be
classified into finite volume, discreteness, and
edge effects (\cite{sc96}). They arise respectively from the (hypothetical)
fluctuations on scales larger than the simulation, the finite
number of particles used to model the density field, and the
uneven weighting of points. 
Due to the periodic boundary conditions,
edge effects are not significant; neither are discreteness
effects except for the smallest scales because of the large
number or particles used. Therefore finite volume effects are
expected to be the dominant contribution,
if systematic errors from the inaccuracy of the calculations
are not considered.

Two methods were used for estimating the cosmic errors: measuring the
dispersion numerically
from the ensemble of simulations $i_{c1}\ldots i_{c5}$
(see the errorbars in Figure 3), 
and using the theory
of (\cite{sc96,scb98}) to estimate the errors 
from the measured higher order moments self-consistently up to 4th order.
The details of the calculations can be found there; here we only
summarize the basic idea.

\cite{sc96} calculated the generating function of the variance
of factorial moments due to edge, discreteness, and finite volume
effects. Since the connected moments can be expressed in terms
of the factorial moments (which are the discrete version of
the disconnected moments), their results can be used to
express the errors on the connected moments (see
\cite{scb98} for more details). The resulting formulae
express the errors on the $N$-th order connected moments
in terms of the $2N$-th connected moments for $N \le 4$.
The expressions  are too complicated to quote here
(they are over $500$ lines long); therefore, only 
the self-consistent numerical estimates are used. 
For the case of the connected moments it is not possible
to simply separate the different contributions for the errors.
Therefore, discreteness and edge effects are included in the calculations, 
even though this way the errors could be
overestimated at large scales according to the previous considerations.

Figure 6 compares
the numerical estimates of the theoretical error calculation
(see also \cite{csvirgo98})
The solid line displays the unbiased
estimate of the variance for $S_3$, and $S_4$, while the
dotted lines show the theoretical calculation of the errors
in the individual simulations $i_{c1}\ldots i_{c5}$. The dashes
are the result of a theoretical calculation as well, but using the
ensemble average
of the five simulations for the moments. While the theoretical
estimates from the individual simulations are in excellent agreement
with the empirical dispersion, the average of the five simulations
curiously overestimates the errors, especially on larger scales.
A possible explanation is that since the error distribution is
skewed (\cite{sc96}), a few overshoots can dominate the average.
This is amplified by the non-linear expressions used to 
estimate the errors. The agreement nevertheless
is surprisingly good, despite the anticipated 'error on the error'
problem (\cite{sc96}): the error on the 4-th order moment depends on 8-th
order quantities, and the error on the error depends on up to
16-th order moments. To determine empirically the errors with
negligible variance, 16 orders should be controlled with high
precision, which is hardly possible using only 5 simulations
of this size. For instance, in \cite{sc96} 1000 subsamples were
needed to control the error on the error. 
Nevertheless, we can draw from the figure the conclusion that
the theoretical calculations for the {\em individual} simulations
are in excellent agreement with the empirical dispersion.
We generalize this finding to the other simulations, where
an ensemble of realizations is not presently available; i.e.
we assume that the theoretical calculation
is a good estimate of the errors up to 4-th order.
The error calculation yields less than $1$ \% error for $S_3$
and about $5$ \% for $S_4$ at most scales, except perhaps
at the largest scales, where the errors appear to turn up
to few tens of \%.

In fact for simulation $I_c$, which has the exact same properties
as simulations $i_{c1}\ldots i_{c5}$ except larger,  it is interesting to try
the following na\^ive scaling: if, as argued above,
finite volume effects dominate, the errors on the disconnected
moments scale with the variance over the full box, $\sqrt{\xi(L)}$.
Even though for the connected moments the formulae are more complicated,
we find empirically that scaling with $2 \sqrt{\xi(L)}$ is
an excellent approximation for the errors. Moreover, it appears
that the earlier outputs have the same absolute error, i.e. the same 
scaling. Encouraged by these
findings, a similar scaling was applied to the $\Omega < 1$ simulations.
Again, scaling the $2 \sqrt{\xi(L)}$ of the last output is
an excellent approximation. These observations are 
valid at the factor of 2 level: a considerable accuracy if the 
arguments about the ``error on the error'' are taken into account.
We conjecture that similar approximations can be used at higher
order. 

\section{Discussion}

The approximations developed for the errors in the previous
section facilitate the comparison of the results with observations.
The framework for comparison is naturally provided by perturbation
theory (PT) and its generalization for smaller scales, extended
perturbation theory (EPT). PT gives simple expressions for the
higher order correlation amplitudes $S_N$ at any order $N$.
For instance for the third order quantity $S_3 = 34/7-(n+3)$
(see, e.g., \cite{jbc93}), where
$n$ is the local index of the power spectrum. This formula, and
the corresponding ones for higher order, can {\em formally} be
used at small scales where PT is not expected to hold. It
was observed in scale invariant $N$-body simulations (\cite{cbh96,cbbh96})
and observations (\cite{smn96}),
that this formal procedure gives an excellent fit for the higher
order cumulants, even though the resulting $\neff$ is no longer
the local slope of the power spectrum; rather, it is a formal parameter
which proves to be extremely useful for characterizing data. 
In scale invariant simulations it was found that a steepening occurs
in terms of $\neff$, i.e. the distribution in terms of its cumulants
at non-linear scales is equivalent to another weakly non-linear
distribution but with a steeper power spectrum. 

The solid lines in Figure 8 show the least square fit for
$\neff$ in all the large simulations. Up to sixth order quantities
were used, and the errorbars were obtained formally by calculating
$\neff$ from $S_3$ alone. This takes into account the inaccuracy of the
higher order moments relative to the third order moments, as
well as the possible variance in EPT, an approximate
phenomenological relation. This prescription, 
however, cannot account for any absolute errors on the measurements of
the $S_N$'s. Figure 8, the most sensitive summary of the results
of the paper, shows $\neff$ on a {\em linear} scale.

The upper three  solid lines correspond to $I_a, I_b, I_c$,
in increasing order on scales of $2\mpc$, 
the lower three to  $II_a, II_b, II_c$, in decreasing order on scales
of $15\mpc$.
The $\Omega = 1$, and $\Omega < 1$
simulation groups are tightly together, while the two groups differ
from each other.  In the weakly non-linear regime the agreement 
is excellent between PT theory and the measured $S_N$'s, 
since the $\neff$ is extremely close to the theoretical
slope of the power spectrum: the upper dashed line shows the theoretical
prediction for $\Omega = 1$, and the lower three dashed lines show
the prediction for $\Omega < 1$, for $II_a, II_b, II_c$ in decreasing
order. Note that the actual $\Omega$ dependence of the $S_N$'s,
which is extremely small, (\cite{bjcp92}) was not taken into account
for the theoretical prediction; simply the local slope
of the power spectrum is plotted. 
 Since the power spectrum is slightly different for 
the $\Omega < 1$ simulations, they behave
differently in this regime. 
Because of non-linearities at smaller scales, PT is not a good approximation;
however, EPT still is. This can be seen from Figure 4, where triangles
show the $S_N$'s formally corresponding to the fitted $\neff$.
The agreement is excellent above $l_{50}$ on all figures, except 
perhaps for $II_c$ where it is only a good approximation above
$2 l_{50}$ for the higher orders.
Thus $\neff$ at each scale
is good representation of the data, providing a natural framework
for comparison. In the highly non-linear regime,  a steepening 
compared to the PT value is present, which is apparent relative to the dashed
line on the Figure. This is very similar to the effect observed
in scale invariant simulations by \cite{cbh96}. Note also that as
the simulations become more relaxed, EPT becomes more
accurate. This suggests that the break down at small scales is
caused only by inaccuracies introduced by dynamical discreteness effects
at small scales. In Figure 8 the same effect shows up
as a fan-like spreading of the curves, corresponding to a
a slight decrease of the $S_N$'s as a function of time, as discussed before. 
When particle discreteness is accounted for,  the $S_N$'s appear to be
approximately time-independent to a degree similar to the weakly
non-linear regime even at highly non-linear scales. 
On the other hand, the difference between the 
two types of simulations is real, as it is observed at the reliable
scales. Note that  by construction the errorbars of the Figure
cannot reflect systematics from particle discreteness.

The framework provided by $\neff$ is ideal for  comparison
with observations. The same type of calculation was performed by
\cite{smn96} using the EDSGC survey. Their results agree well with 
the corresponding $S_N$'s from the APM survey (\cite{gaz94,sg98}).
The $\neff$ from the EDSGC is plotted with dotted lines.
Although the split between the different
output times could be artificial, as noted above, the difference
between the $\Omega = 1$ and $\Omega < 1$ simulations is
real. The comparison with the EDSGC data clearly favors
the $\Omega < 1$ curves.
Note a subtlety of the comparison shown here: the scales given
with the deprojected $S_N$'s in \cite{smn96} are simply $D\theta$,
where $D = 370\mpc$ is the depth of the catalog, and
$\theta$ is the angle of the sides of the square window used
for counts in cell.
Since the simulation uses cubical cells, the comparison with $D\theta$
is not appropriate. On the figure a simple approximation is used:
the volume of the effective cone (or pyramid) is equated to the
volume of the cells in the simulations. More precisely,
$D^3\theta^2/3 = l^3$ was assumed, where $l$ is the side of the
cubes in the simulations. If $\theta_{deg}$ is expressed in degrees, the scale
transformation is $l = 35.9 \theta_{deg}^{2/3}\mpc$ as opposed to the usual
$l = 6.5 \theta_{deg}\mpc$. Comparing the
volumes should be a reasonable approximation on small
scales, where virialization erases any configuration dependence,
but it is expected to break down on larger, weakly non-linear scales, where
the elongated pyramids might have different $S_N$'s than the
equivalent cubes (\cite{ssf98}). 

While it is clear from the  Figure that the data favor the $\Omega < 1$
models, let us use a toy model of biasing to quantify this statement.
This should be reasonably accurate in the weakly non-linear
regime, even though the configuration dependence of the higher
order moments start to enter the picture.
Here we use the ansatz $S_3 = 34/7 - (\neff +3)$ from EPT,
and the leading order bias formula $S_3^g = S_3/b+3 b_2/b^2$
(\cite{fg93}), where the galaxy field is expanded in a Taylor
series as $\delta^g = b \delta + b_2 \delta^2/2 + \ldots$, 
and the superscript $g$ signifies galaxies.
If one formally applies EPT for the (possibly) biased galaxy field,
as was done in the case of EDSGC (\cite{smn96}), it is possible
to express $b_2$ in terms of the measured effective indices
of the galaxies $n_g$ as
\begin{equation}
  b_2 = - \frac{b}{21}(13-13b - 7\neff  + 7 b n_g).
\end{equation}
In this equation $b$ is fixed by the $\sigma_8$ of the simulation,
$\neff$ and $n_g$ are the measured effective index in the
simulation, and in the galaxy catalog,  respectively.
Note that the two-point functions of the different time
outputs differ essentially only in the amplitude within the
studied scale range, thus bias is approximately scaleindependent,
and can be described by $\sigma_8$.
The results of such a model are shown in Figure 9.
$b_2$ is plotted against scale from $4-20\mpc$. 
Note that the two outlying points above $20\mpc$ on Figure 8.
are caused by edge effects in the EDSGC survey \cite{smn96}.
The curves in increasing order represent simulations
$II_c, II_b, I_c, II_a, I_b, I_a$. The interpretation of the
Figure is not straightforward, since the leading order calculations
are only expected to work on large scales, where
the errors of the EDSGC measurements and the configuration
dependence are becoming increasingly influential.
Nevertheless, it is intriguing that the $\Omega = 0.3, b = 1$
($II_c$) model requires no nonlinear biasing in terms of $b_2$ within
the errors,
in contrast with all the other models. Thus it is possible to explain
the higher order distribution of counts in cells in the EDSGC without
invoking linear or non-linear biasing. The minimal assumption
that the EDSGC galaxies trace mass satisfies the above model, but none
of the others examined. Occam's razor rejects them as they 
all need significant non-linear biasing. 

Note, however, Figure 9 is not the ultimate answer.
More investigations of the non-linear bias are needed, where
the above simple theory breaks down, because of the non-linearities
and stochasticity. Also, the configuration dependence can be modeled
more accurately by the use of artificial catalogs with realistic
selection functions, which employ pyramid shape cells as 
used in the EDSGC. In addition, more data, especially with redshifts, with larger
dynamic range towards large scales will turn this argument into a more
quantitative result. The possible extensions are left for future
work, while the data requirements will be met by the new generation
of galaxy surveys.
In the near future, the Sloan Digital Sky Survey, and the 2 degree field
Survey will determine the higher order moments with 
similar accuracy to the present simulations. Comparison of the future
data with the results reported here will strongly constrain biasing models.

This paper presented the measurements of cumulants 
of counts in cells in
CDM and $\Omega$CDM simulations. These high resolution simulations
together with a pair of new measurement algorithms
enabled us to explore a larger dynamic range with smaller
errors than previously was possible. A careful attempt was
made to determine the range of reliable scales, and
a fully non-linear theoretical error calculation was performed as well.
It was found that, via perturbation theory and
extended perturbation theory, the results can be efficiently
represented by the effective index, $\neff$.
In the weakly non-linear regime excellent agreement was found
with perturbation theory, while at smaller scales a non-linear
plateau found in scale invariant simulations was confirmed. 
At small scales, the agreement with extended perturbation theory
was found to be remarkably good. 
The CDM results are qualitatively similar
to the scale invariant simulations in all respect considered.
The time dependency of the
cumulants appears to be negligible at all scales, if
particle discreteness is correctly taken into account.
A comparison with observations revealed that the $\Omega < 1$
model is consistent with the higher order correlations of
the EDSGC galaxies without the need of biasing. The rest
of the models examined need substantial non-linear biasing 
to be reconciled with the data.
Finally, the error formulae of
\cite{sc96,scb98} provide a good approximation for the errors
on higher order statistics measured in $N$-body simulations.

I.S. would like to thank Stephane Colombi for stimulating discussions,
Carlton Baugh and Shaun Cole for useful suggestions, and 
Albert Stebbins for instrumental help.
I.S. was supported by the PPARC rolling grant for 
Extragalactic Astronomy and Cosmology at Durham. 
This work was also
supported by DOE and NASA through grant NAG-5-2788 at Fermilab, and 
by a NASA HPCC-ESS grant.  The simulations were carried out
on the Pittsburgh Cray T3E, the Goddard Space Flight Center Cray T3E,
and the Cray T3D at the Arctic Regional Supercomputing Center.

\section{Appendix}

Here we discuss the algorithms used to calculate counts in cells.
While from theoretical point of view it would be ideal to use
the infinitely oversampling algorithm of \cite{sz97}
to estimate the distribution of counts in cells (\cite{sc96}),
this would be unrealisticly slow even with present day computers
for 47 million particles in three dimension. The algorithms discussed here
provide an efficient way to sample with $\simeq 10^{9-14}$ cells
in less than 8 CPU hours on a typical workstation with approximately
1GB of memory. The two methods are complimentary to each other, 
one for large, the other for small scales, 
with ample overlap between the two. They are outlined next.

\subsection{Large Scales (A1)}

The algorithm for large scales will be 
explained in one dimension for simplicity. The generalization for arbitrary
dimensions is obvious. The three dimensional version was
used in the calculations of this paper.

The computations are performed on the largest possible grid 
with $N$ segments
which can be fit into the memory of the computer: this determines
the smallest possible scale $L/N$, where $L$ is the box size,
and $N$ is the base sampling.
A hierarchy of scales are used, with the scale at a given level being
twice the scale at one level lower.
The results one step lower in the hierarchy are used to keep
the number of sampling cells constant even at the largest scales.
Counts in cells can be straightforwardly calculated on the resolution 
scale of the grid, i.e. the smallest scale considered. For the calculation
at twice the previous scale the sum of two cells
are always stored in one of the cells, for instance in the one with smaller
index.  Because of the
periodic boundary conditions, auxiliary storage is required to 
calculate the sum of the values in the rightmost cell (if the
summations was done left to right), as its right
neighbor is the leftmost cell which was overwritten in the first step.
After these preparatory steps
counts in cells can again be calculated from the $N$ numbers 
representing partially overlapping cells. For the next level, twice the
previous scale, one needs the sum of four original resolution cells:
a calculation simply done by summing every other cell of the previous
results into one cell. 
At this level, two auxiliary storage spaces are needed because of the
periodicity. In general, at each level in the hierarchy
two cells of the previous results are summed as 
a preparatory step, and counts in cells are calculated simply
by computing the frequency distribution of the $N$ sums stored in
the main grid. Auxiliary storage is needed for those rightmost cells,
which have the periodic neighbors on the left end.

In D dimensions $2^D$ cells are summed in the preparatory step,
and the auxiliary storage space enlarges the original hypercube.
In our case the main grid was a $512^3$, 
resulting in a $1\mpc$ spacing of $1.3 \times 10^8$ cells.
Further precision could be achieved by oversampling the 
original grid, that is, shifting it by a fraction of
a resolution cell. Our CPU resources allowed for one independent
shifting in each direction, which resulted in an 8 times oversampling
of the original grid, that is $C = 1.1 \times 10^{9}$ cells at each scale
from $\simeq1\mpc$ to $\simeq128\mpc$, i.e. a quarter of the length of the box.

\subsection{Small Scales (A2)}

The above procedure is limited at small scales 
by the largest grid that will fit into the memory of the computer.
Therefore an alternative technique was adapted for small scales
using the original oct-tree data-structure 
of tree $N$-body codes.
This is an efficient representation of a 
sparse array, since at small scales most of the cells are empty
in a grid spanning the simulation.
 The tree is built up recursively, by always dividing
the particles into two groups based on which half of the volume
they belong to. The same function is called on both halves
with the corresponding particles until
there is no particle in the volume, or the scale becomes smaller
than a predetermined value. At each level the scale and the number
of particles are known, and when an empty volume is reached, all
contained volumes are also empty. These two observations are enough
to insert the book-keeping needed to calculate counts in cells
at all scales while the tree is built. The number of sampling
cells at each level are $2^l$, where $l$ is the level; the original
box is  represented by $l = 0$. Towards smaller scales the
number of cells increases. When $N^3 = 2^l$, where $N$ is the
size of the largest grid of the previous algorithm, the two
techniques should (and do) give the exact same answers. At larger
scales the previous algorithm is superior, since $N > 2^l$,
while this algorithm becomes useful at smaller scales. Just as above,
this procedure can be further improved by shifting the particles
slightly before calculating the tree. However, since this hierarchy
of grids has different numbers of cells, random shifts are more
advantageous. Shifting by a fraction of the smallest scale would not
exhaust the possibilities for any larger scale, while shifting by
a fraction of the largest grid might not shift the underlying grids
at all. 
With the introduction of random shifts (oversampling grids),
the dynamic range of the two algorithms develops a substantial
overlap, which is useful for testing. According to Figure 3. the
algorithms produced essentially the same higher order moments
in the overlap range of five twofolds. 

\hfill\newpage

\section{Figure Captions}

\noindent  Figure 1. Counts in cells distribution for simulation
$I_c$ (see text) from scales of $7.63\kpc$ doubling up to $125\mpc$
from left to right. The scale corresponds to the size of the cubical
window. Algorithm A1 was used for scales $ \ge 1\mpc$, and A2 was used for
the smaller
scales. Note that the accuracy is as much as $10^{-14}$, and at least
$10^{-9}$.

\noindent Figure 2. Upper Panel: The joining solid and dashed lines display
the measured variance $\xiav$ in cubical cells as a function of
scale as calculated with algorithms A1 (solid) and A2 (dashed).
The triangles and squares are the predictions from the linear and
nonlinear power spectra, respectively. The dotted line is
the same for simulation $i_0$. Lower Panel: $N_c = \xiav \bar N$
as a function of scale. The line types are the same as for the 
upper panel.

\noindent Figure 3. The cumulants $S_N$  are displayed in increasing
order upwards up to 9th order as measured
in simulation $I_c$ with algorithms A1 (solid), and A2 (dashed).
The dot-dashed lines show the predictions of perturbation theory
up to 6th order. The dotted lines are similar measurements in
simulation $i_0$. The triangles with errorbars display the
the average and dispersion of measurements  in simulations 
$i_1\ldots i_5$. Only the upper errorbar is plotted
to reduce clutter.

\noindent Figure 4. The measurements of $S_N$, $3 \le N \le 10$,
in all simulations are plotted with solid lines, in increasing
order upwards. The triangles represent the $S_N$'s
from extended perturbation theory, when the best fit $\neff$
is used.

\noindent Figure 5. The relative measurement errors are plotted
as a function of order. For $N \le 5$ the errors were calculated
with the formulae of \cite{sc96}. The dotted line is a conservative
extrapolation of the calculations, which is likely to overestimate
the actual errors.
 
\noindent Figure 6. The absolute errors for $S_3$ and
$S_4$ are plotted for simulations $i_1\ldots i_5$. The lower set
of curves represent $S_3$, the upper set $S_4$. The solid lines
show the errors estimated by calculating the variance in the
fives simulations. The dotted lines show the theoretical 
error for each of the five simulations calculated from the 
measured cumulants self-consistently.  The dashed lines are the errors
estimated from moments determined by the ensemble average of the five
simulations.

\noindent Figure 7. The theoretical errors for $S_3$ and $S_4$
are shown in simulations $I_a,I_b,I_c,II_a,II_b,II_c$. The lower
set of curves refer to $S_3$, the upper set to $S_4$. Within
each set there are two solid lines with approximately matching
dotted lines. The higher quadruplet refers to simulations
$II_*$, while the lower one to $I_*$. The solid lines show
the dispersion measured in simulations $i_1\ldots i_5$ scaled
with $2 \sqrt(\xiav)$ of $I_c$ (lower), and $II_c$ (higher),  respectively.
The three dotted lines in each quadruplet display the different
output times $a,b,c$.

\noindent Figure 8. The measurements of the higher order moments
are summarized in terms of $\neff$ of EPT. The upper three
solid lines with errorbars display the fitted $\neff$ for simulations
$I_a,I_b,I_c$, increasing in this order on scales of $2\mpc$.
The lower three solid lines correspond to $II_a,II_b,II_c$,
decreasing in this order on scales of $15\mpc$. The upper dashed
line is the theoretical local slope of the power spectrum
for the $\Omega = 1$ simulations, the lower three dashed
lines are the same for $II_a,II_b,II_c$, decreasing in this
order. The dotted lines with errorbars show the measurements
of $\neff$ in the EDSGC survey. The errorbars on this figure
were determined by calculating $\neff$ from $S_3$ alone, and comparing it
with a simultaneous fit using $S_3, \ldots, S_6$. The latter
is displayed, while the difference of the two is an indication
of the accuracy.

\noindent Figure 9. The first non-linear coefficient in the
Taylor expansion of the bias, $b_2$, is displayed for the
different simulations as a function of $\log$ scale. 
The curves, in increasing order, are for 
$II_c, II_b, I_c, II_a, I_b, I_a$.

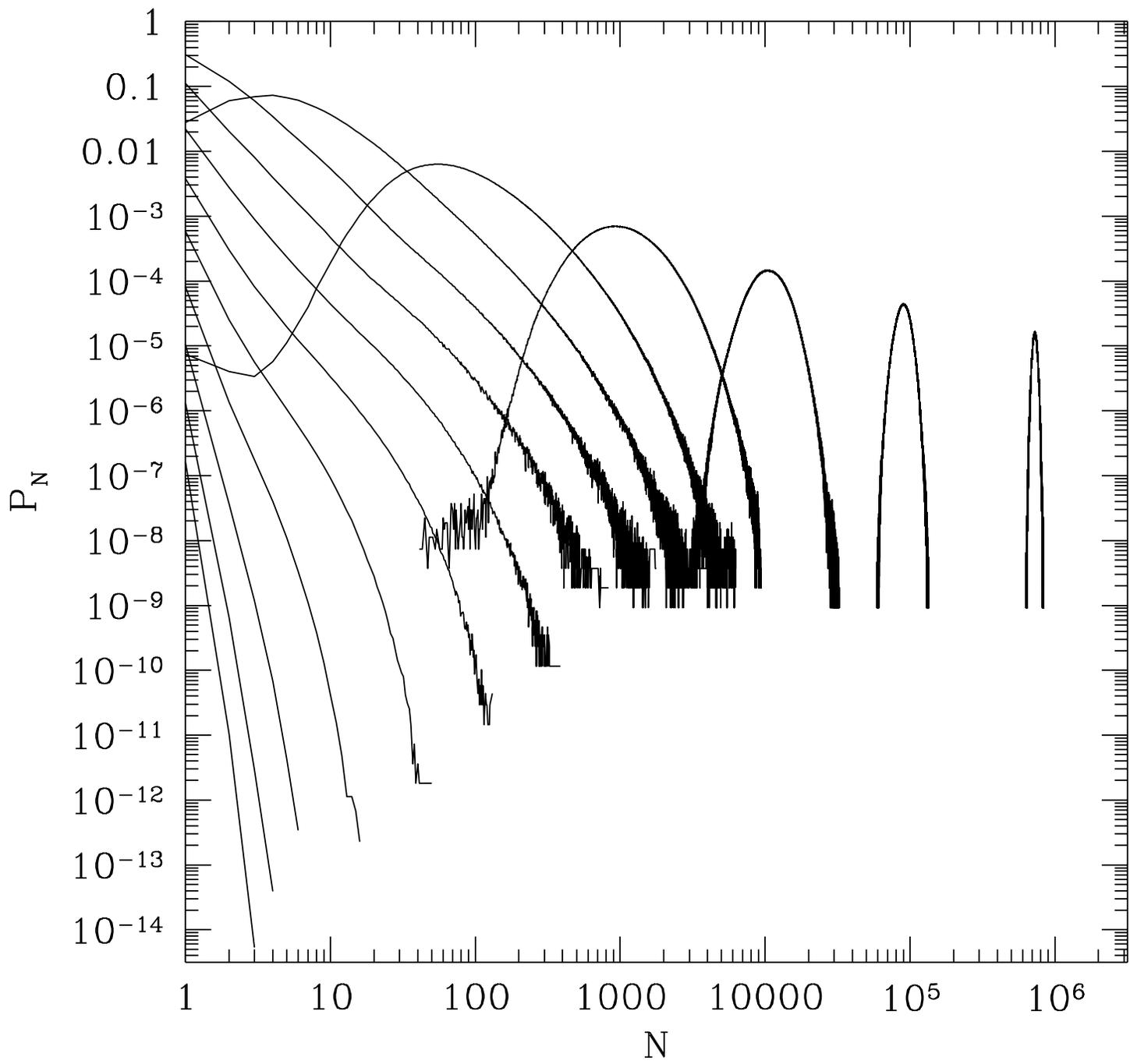
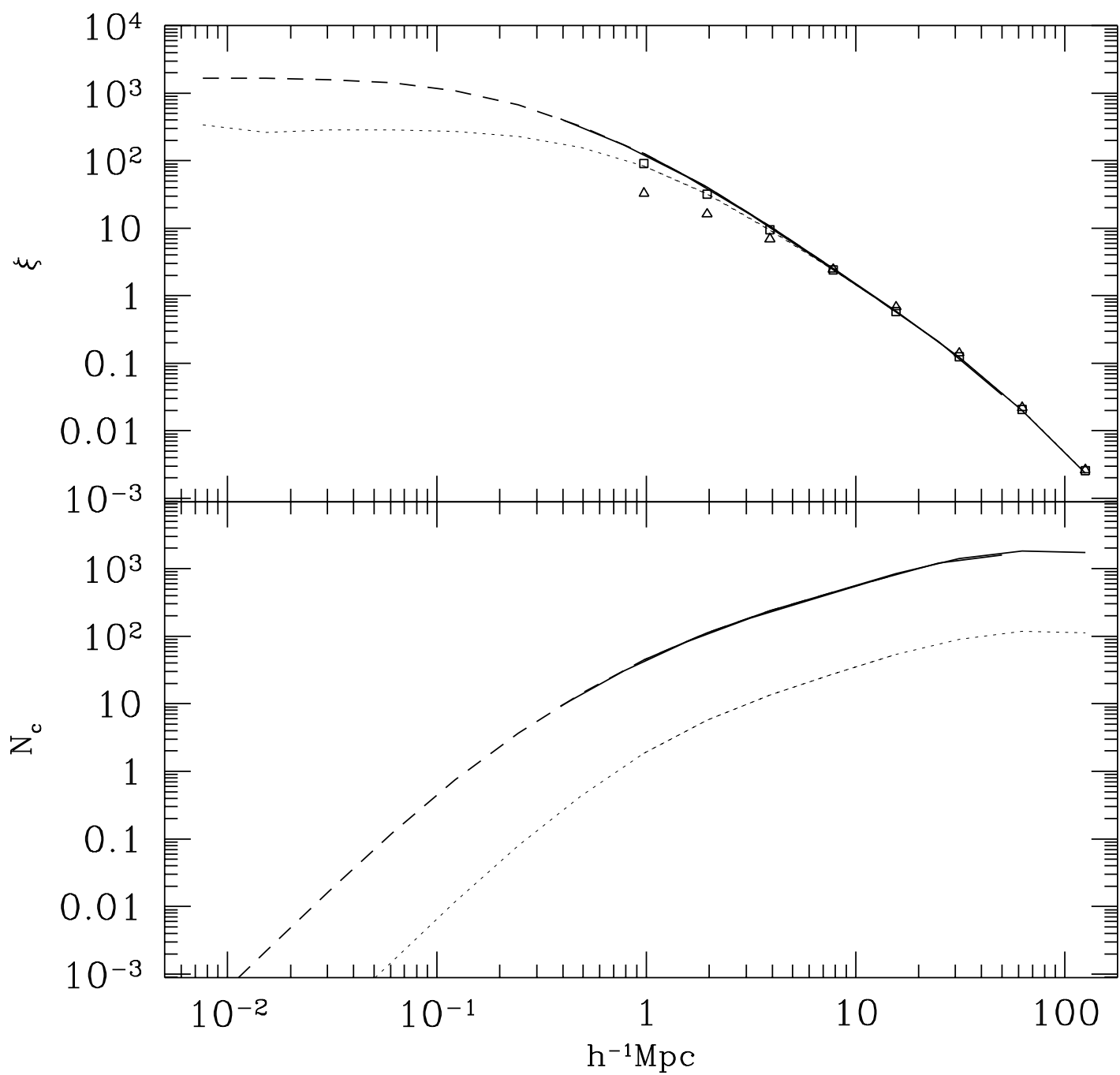
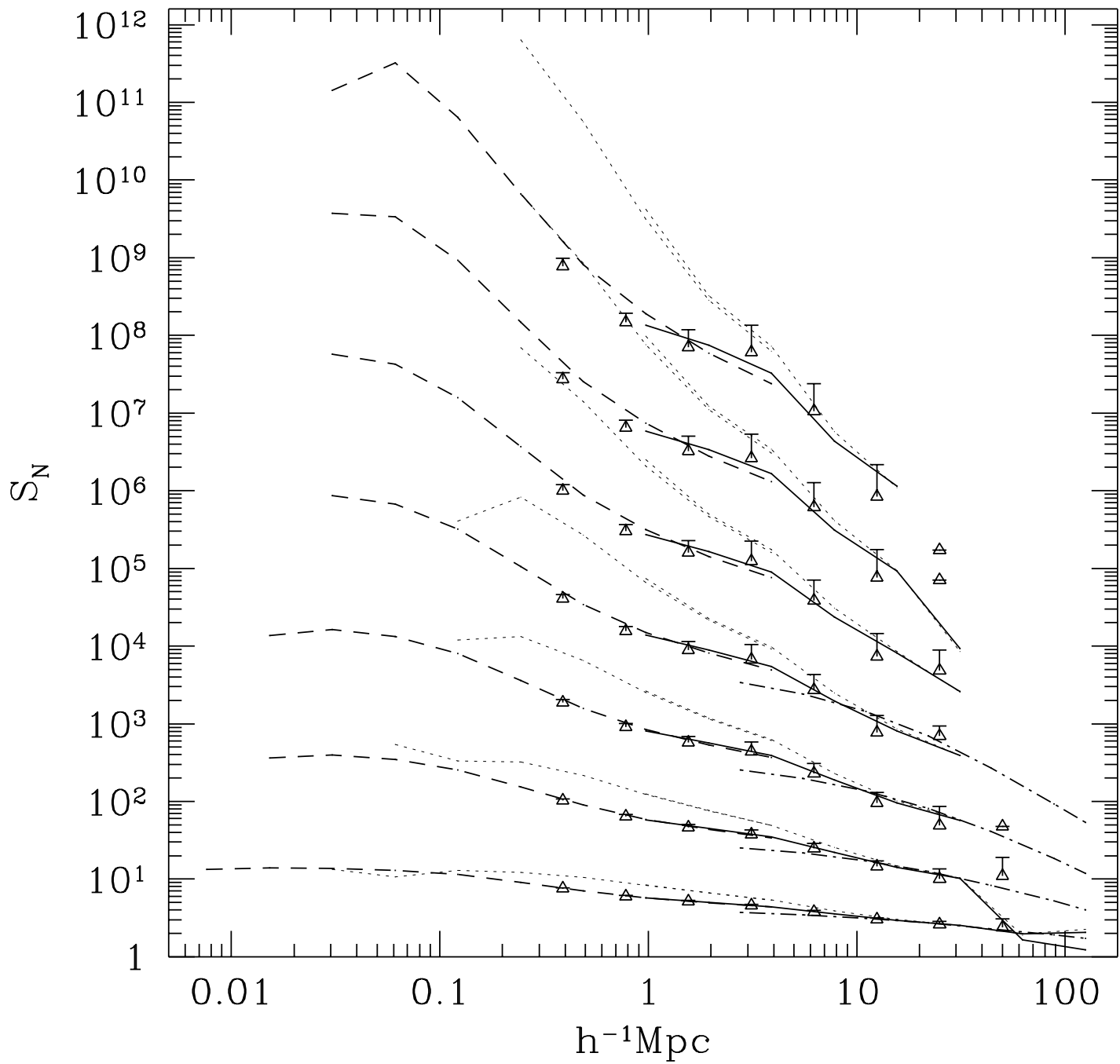
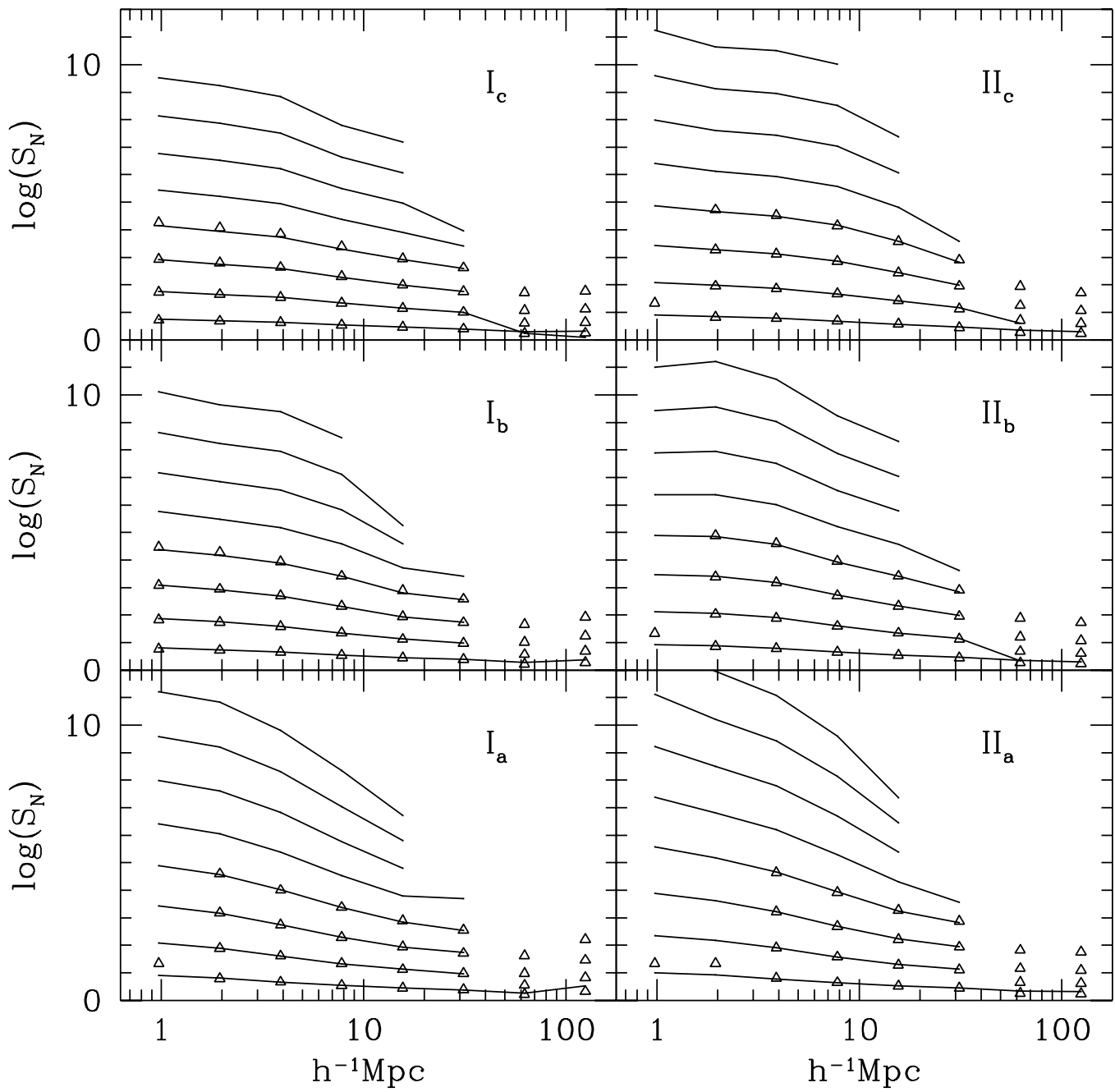
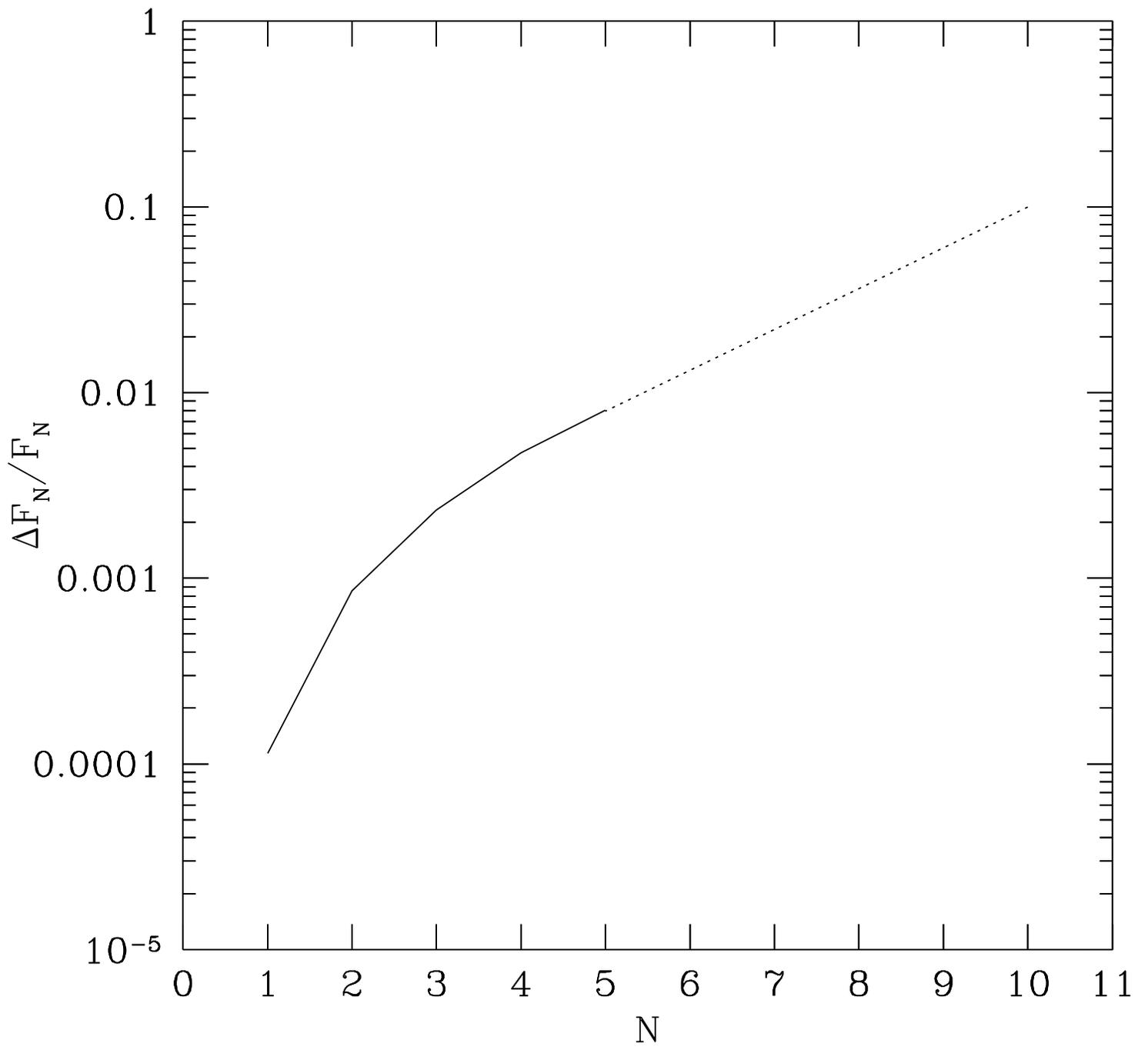
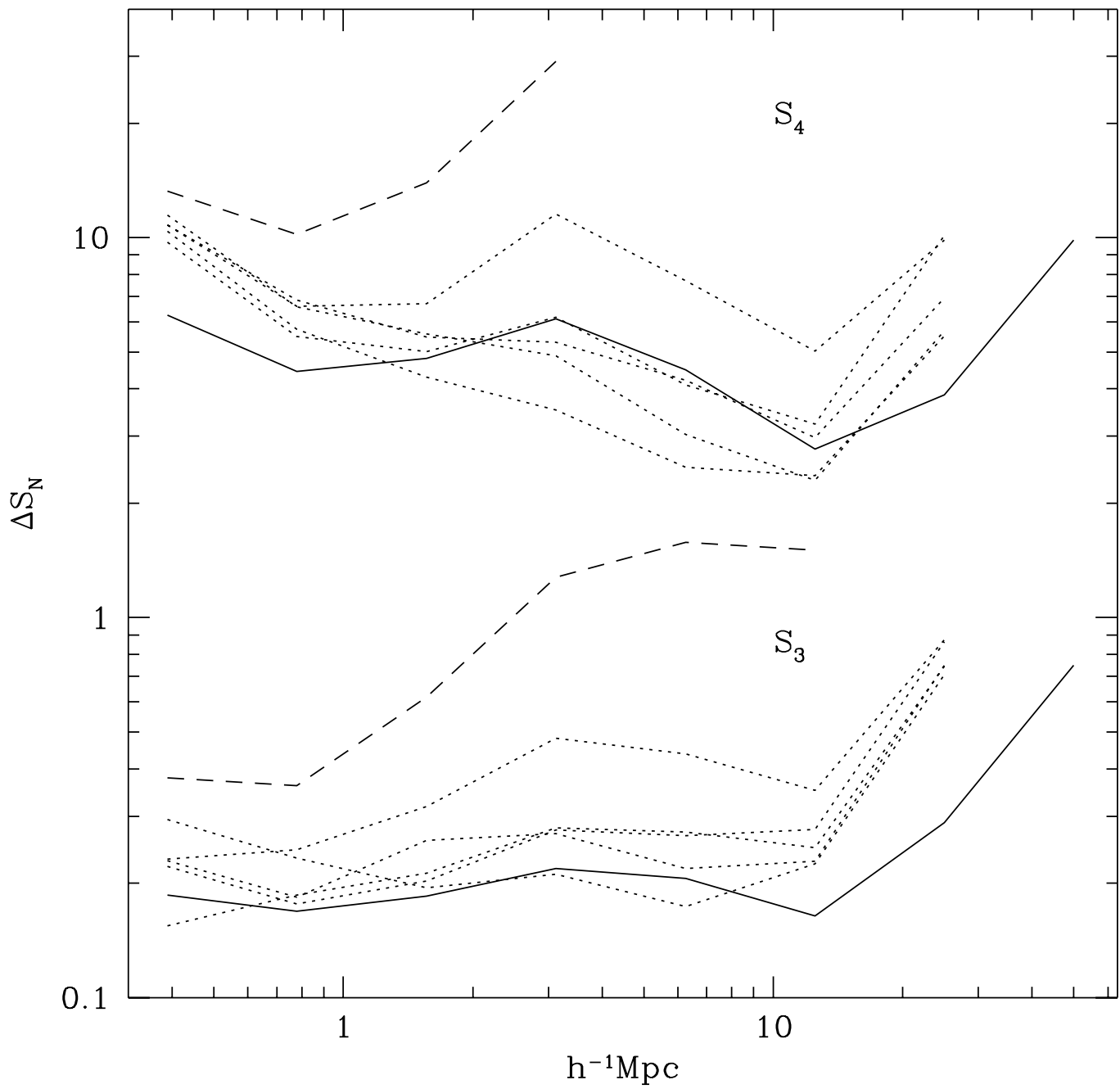
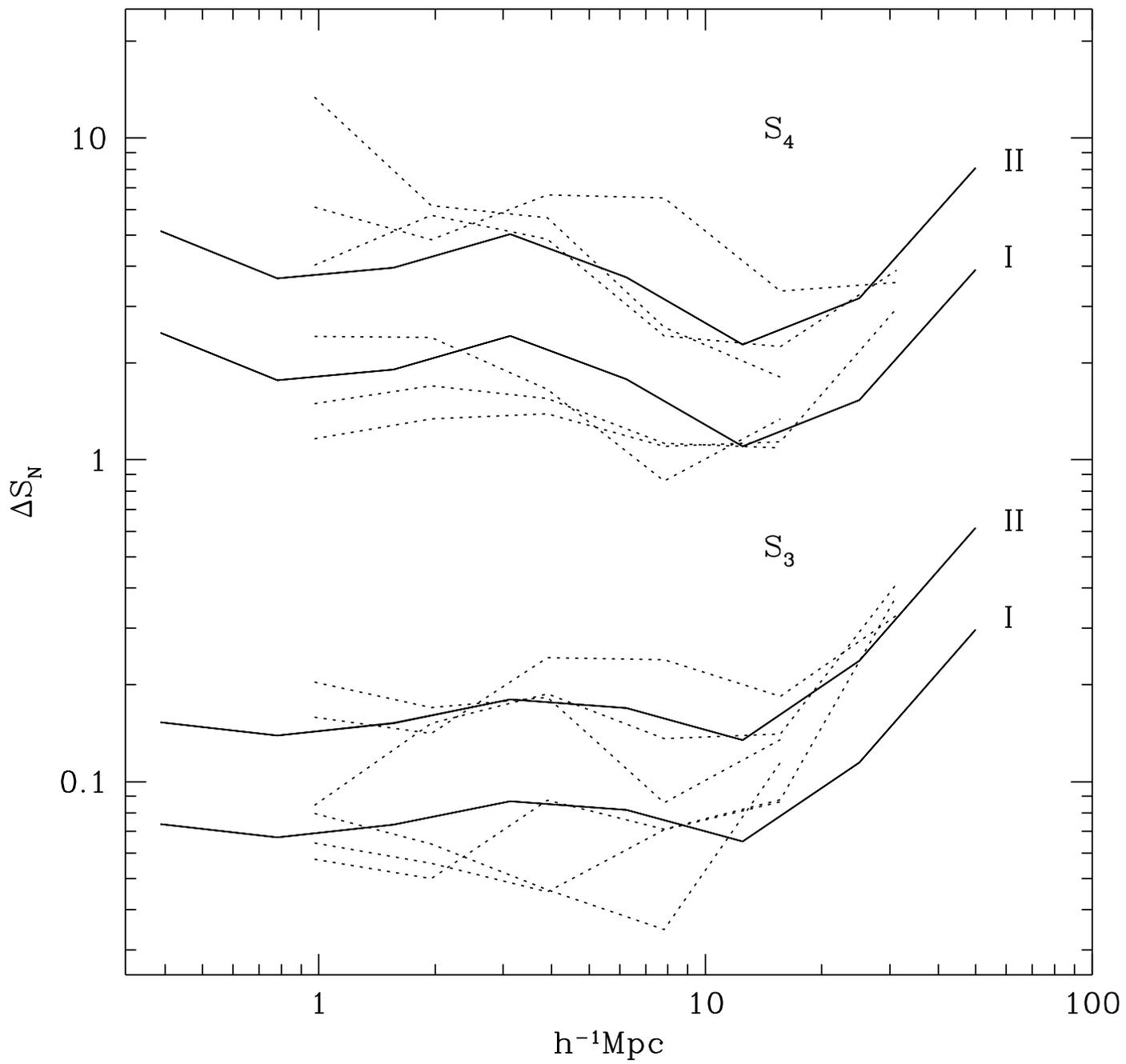
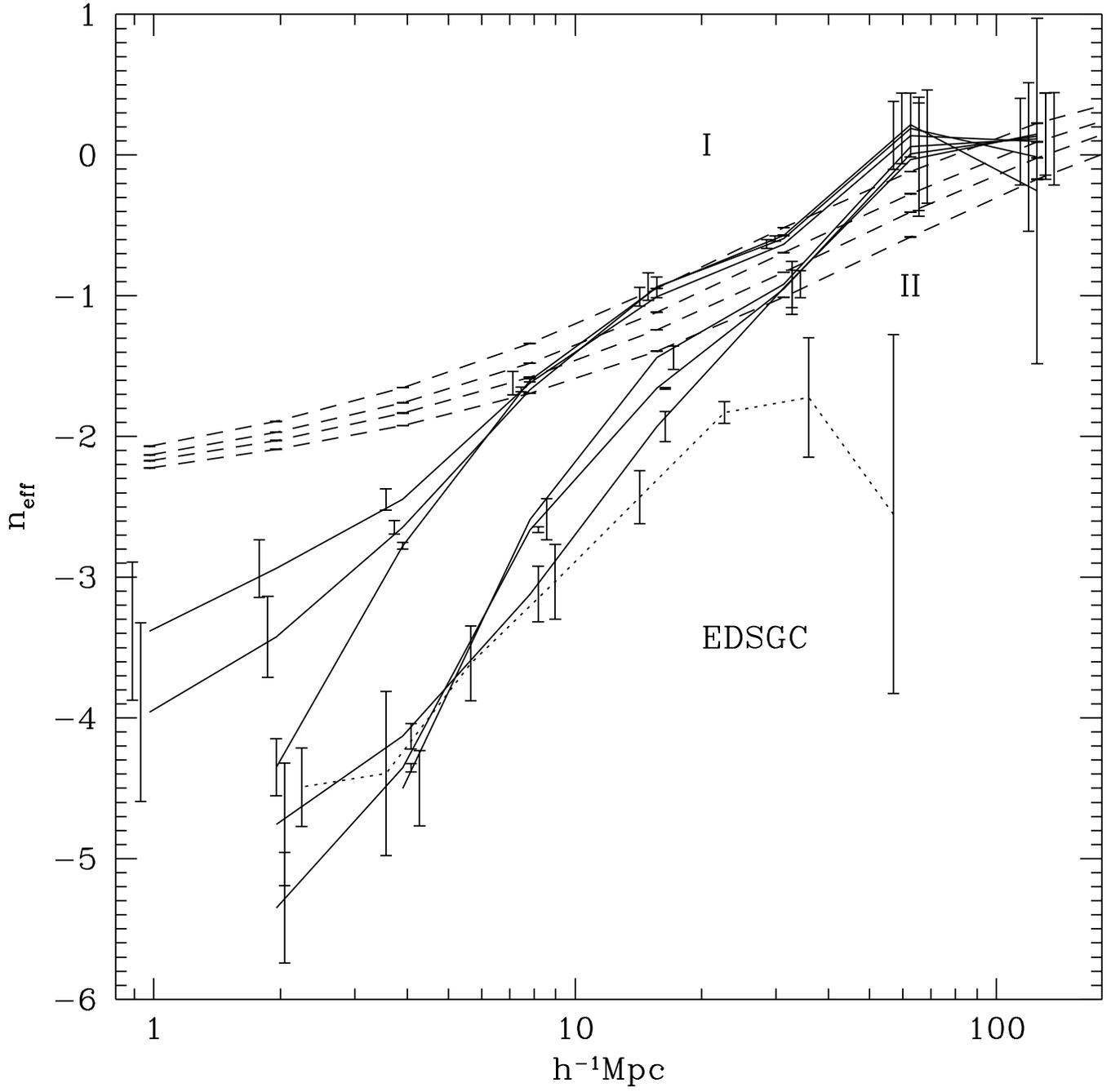
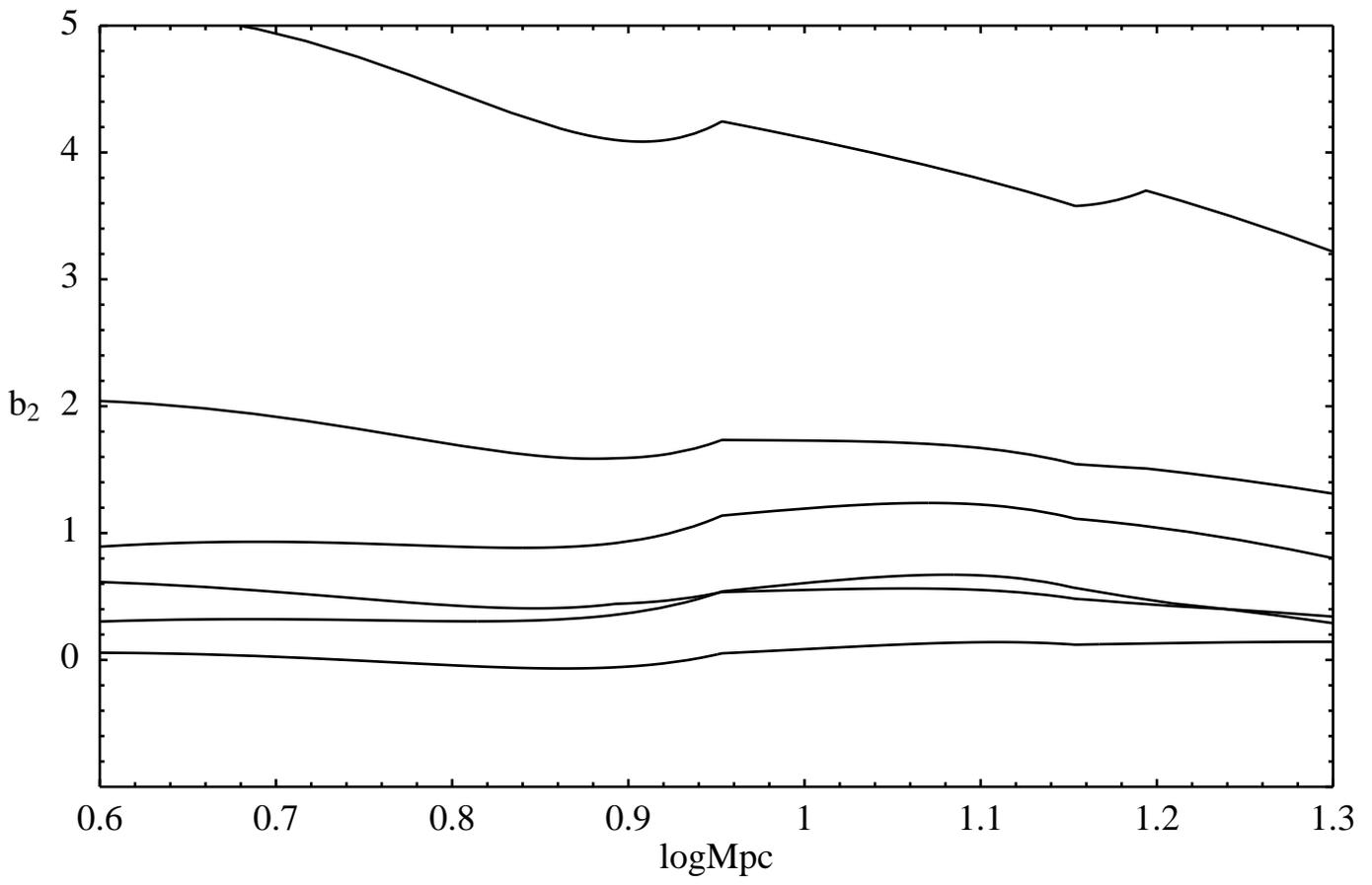
\end{document}